\documentclass[fleqn,10pt]{wlscirep}

\usepackage{graphicx}
\usepackage{subcaption}

\usepackage[T1]{fontenc}
\usepackage[utf8]{inputenc}
\usepackage[scaled]{helvet}
\usepackage[mathletters]{ucs}

\usepackage{textcomp}  

\usepackage{gensymb}   
\begin{document}

\title{Enabling real-time multi-messenger astrophysics discoveries with deep learning}

\author[1,*]{E. A. Huerta}
\author[1]{Gabrielle Allen}
\author[2]{Igor Andreoni}
\author[3]{Javier M. Antelis}
\author[4]{Etienne Bachelet}
\author[2]{G. Bruce Berriman}
\author[5]{Federica B. Bianco}
\author[6]{Rahul Biswas}
\author[1]{Matias Carrasco Kind}
\author[7]{Kyle Chard}
\author[8]{Minsik Cho}
\author[9]{Philip S. Cowperthwaite}
\author[10]{Zachariah B. Etienne}
\author[7]{Maya Fishbach}
\author[11]{Francisco F\"{o}rster}
\author[12]{Daniel George}
\author[13]{Tom Gibbs}
\author[2]{Matthew Graham}
\author[1]{William Gropp}
\author[1]{Robert Gruendl}
\author[1]{Anushri Gupta}
\author[1]{Roland Haas}
\author[1]{Sarah Habib}
\author[14]{Elise Jennings}
\author[1]{Margaret W. G. Johnson}
\author[15]{Erik Katsavounidis}
\author[1]{Daniel S. Katz}
\author[1]{Asad Khan}
\author[1]{Volodymyr Kindratenko}
\author[1]{William T. C. Kramer}
\author[1]{Xin Liu}
\author[2]{Ashish Mahabal}
\author[16]{Zsuzsa Marka}
\author[1]{Kenton McHenry}
\author[17]{J. M. Miller}
\author[18]{Claudia Moreno}
\author[1]{M. S. Neubauer}
\author[13]{Steve Oberlin}
\author[19]{Alexander R. Olivas Jr}
\author[1]{Donald Petravick}
\author[1]{Adam Rebei}
\author[1]{Shawn Rosofsky}
\author[1]{Milton Ruiz}
\author[1]{Aaron Saxton}
\author[20]{Bernard F. Schutz} 
\author[1]{Alex Schwing}
\author[1]{Ed Seidel}
\author[1]{Stuart L. Shapiro}
\author[1]{Hongyu Shen}
\author[1]{Yue Shen}
\author[21]{Leo P. Singer}
\author[22]{Brigitta M. Sip\H{o}cz}
\author[1]{Lunan Sun}
\author[1]{John Towns}
\author[1]{Antonios Tsokaros}
\author[1]{Wei Wei}
\author[23]{Jack Wells}
\author[14]{Timothy J. Williams}
\author[8]{Jinjun Xiong}
\author[1]{Zhizhen Zhao}

\affil[*]{e-mail: elihu@illinois.edu}

\affil[1]{University of Illinois at Urbana-Champaign, Urbana, Illinois 61801, USA}
\affil[2]{California Institute of Technology, 1200 E California Blvd, Pasadena, California 91125, USA}
\affil[3]{Tecnologico de Monterrey, School of Engineering and Sciences, Zapopan, Jalisco 45138, Mexico}
\affil[4]{Las Cumbres Observatory, 6740 Cortona Drive, Suite 102, Goleta, California 93117, USA}
\affil[5]{University of Delaware, Newark, Delaware, 19716, USA}
\affil[6]{The Oskar Klein Centre for Cosmoparticle Physics, Stockholm University, AlbaNova, Stockholm SE-106 91, Sweden}
\affil[7]{The University of Chicago, Chicago, Illinois 60605, USA}
\affil[8]{IBM T.J. Watson Research Center, Yorktown Heights, New York 10598, USA}
\affil[9]{Observatories of the Carnegie Institute for Science, 813 Santa Barbara Street, Pasadena, California 91101-1232, USA}
\affil[10]{West Virginia University, Morgantown, West Virginia 26506, USA}
\affil[11]{Center for Mathematical Modeling, Beaucheff 851, 7th floor, Santiago, Chile}
\affil[12]{Google X, Mountain View, California 94043, USA}
\affil[13]{NVIDIA, 2788 San Tomas Expressway Santa Clara, California, 95050}
\affil[14]{Argonne National Laboratory, Leadership Computing Facility, Lemont, Illinois 60439, USA}
\affil[15]{Massachusetts Institute of Technology, 77 Massachusetts Ave, Cambridge, MA 02139, USA}
\affil[16]{Columbia University in the City of New York, New York 10027, USA}
\affil[17]{Center for Nonlinear Studies, Los Alamos National Laboratory, Los Alamos, New Mexico 87545, USA}
\affil[18]{Universidad de Guadalajara, Centro Universitario de Ciencias Exactas e Ingenieria, Guadalajara, Jalisco 44430, Mexico}
\affil[19]{University of Maryland, College Park, Maryland 80742, USA}
\affil[20]{Cardiff University, Cardiff CF24 3AA, United Kingdom}
\affil[21]{NASA Goddard Space Flight Center, Code 661, 8800 Greenbelt Rd., Greenbelt, MD 20771, USA}
\affil[22]{University of Washington, 3910 15th Avenue NE, Seattle, Washington 98195, USA}
\affil[23]{Oak Ridge National Laboratory, Oak Ridge, TN 37831}

\begin{abstract}
\noindent Multi-messenger astrophysics is a fast-growing, interdisciplinary field that combines
data, which vary in volume and speed of data processing, from many different instruments that
probe the Universe using different cosmic messengers: electromagnetic waves, cosmic rays,
gravitational waves and neutrinos. In this Expert Recommendation, we review the key challenges
of real-time observations of gravitational wave sources and their electromagnetic and astroparticle
counterparts, and make a number of recommendations to maximize their potential for scientific
discovery. These recommendations refer to the design of scalable and computationally efficient
machine learning algorithms; the cyber-infrastructure to numerically simulate astrophysical
sources, and to process and interpret multi-messenger astrophysics data; the management of
gravitational wave detections to trigger real-time alerts for electromagnetic and astroparticle
follow-ups; a vision to harness future developments of machine learning and cyber-infrastructure
resources to cope with the big-data requirements; and the need to build a community of experts
to realize the goals of multi-messenger astrophysics.

\end{abstract}
%\begin{document}

\flushbottom
\maketitle

\vspace{4mm}
\noindent Within two years of operation (2015-2017), the advanced Laser Interferometer Gravitational-Wave Observatory (LIGO) and Virgo detectors made unprecedented discoveries, including several gravitational wave (GW) observations of binary black hole (BBH) mergers~\cite{o1o2catalog}. These advances, recognized with the 2017 Nobel Prize in  Physics, have firmly established gravitational wave astrophysics as a field, adding GWs to the existing arsenal of cosmic messengers, namely, neutrinos, cosmic rays and electromagnetic waves. The combination of these complementary signals, known as Multi-Messenger Astrophysics (MMA), have enabled groundbreaking discoveries, such as the observation of supernovae 1987A with neutrinos and electromagnetic waves~\cite{Arnet_Bahcall_1989}, the binary neutron star (BNS) merger GW170817 with gravitational and electromagnetic waves~\cite{bnsdet:2017,2017arXiv171005836T}, and the observation of the blazar TXS 0506+056 with neutrinos and $\gamma$-rays~\cite{blazar_147}.

As the worldwide network of kilometer-scale GW detectors continues to expand, and each detector gradually reaches design sensitivity, the volume of space they probe will continue to grow, increasing the number of GW observations in upcoming observing runs to \(110-3840\,\textrm{Gpc}^{-3} \textrm{yr}^{-1}\) for BNS, and \(9.7-101\,\textrm{Gpc}^{-3} \textrm{yr}^{-1}\) for BBH, at the 90\% confidence level. Upper limits at the 90\% confidence level for the neutron star-black hole (NSBH) merger rate is \(610\,\textrm{Gpc}^{-3} \textrm{yr}^{-1}\)(REF.\cite{o1o2catalog}). Furthermore, next-generation electromagnetic surveys will significantly increase the survey area, field of view, alert production of new and unexpected astronomical events, volumes of image data and catalog sizes, and computational resources to process the image data and prepare annual data releases~\cite{LSST_numbers}. For instance, the Large Synoptic Survey Telescope (LSST)\cite{lsstbook} will collect about 20 TB of data within a 24 hr period. Its alert latency will be 60 seconds, producing up to 10 million alerts per night. With a field of view of 9.6 \(\textrm{deg}^2\), it will survey a total area of 18,000 \(\textrm{deg}^2\), and will complete about 1000 visits per night. By the end of the survey, the final raw image data will be 60 PB, requiring 0.4 exabytes of disk storage~\cite{LSST_numbers,2019LSSTNat}. The convergence of all-sky GW observations with deep, high-cadence electromagnetic observations with next-generation astronomical facilities will provide unique opportunities for new discoveries. 

To realize the full potential of MMA, the computational challenges that currently limit the scope of existing
multi- messenger searches need to be addressed. First, the computational cost of low-latency GW searches based
on implementations of matched- filtering (an optimal, linear, signal-processing tool to detect a signal of known
or expected shape in the presence of additive noise~\cite{owen:1999PhRvD..60b2002O}) 
is such that they can only probe a 4D signal manifold (parametrized by the masses of the binary components  \((m_1,\,m_2)\), and a configuration in which the 3D spin vectors of the objects, \((\mathbf{s}_1,\,\mathbf{s}_2)\), are aligned or anti-aligned with the orbital angular momentum, \(\mathbf{L}\), of the binary system, that is, the only non-zero components of the 3D spin vectors are \((s_1^z,\,s_2^z))\) out of the 9D parameter space available to GW detectors ((\(m_1,\,m_2,\,\mathbf{s}_1,\,\mathbf{s}_2)\) and the orbital eccentricity \(e\)). Extending these algorithms to sample higher-dimensional signal manifolds is computationally unfeasible, since this would require the use of data sets of modeled waveforms at least an order of magnitude larger than existing ones~\cite{Huerta:2017a,2016PhRvD..94b4012H}, which already require the use of supercomputers and the Open Science Grid for core GW data analyses~\cite{HuertaBWOSG,HuertaES,2017Weitzel}. Furthermore, algorithms that search for time-correlated transients in multiple detectors, making minimal assumptions about the morphology of GW signals~\cite{D6:2016}, are tailored for burst-like GW sources, and they may miss second-long GW signals with moderate signal-to-noise ratios. 

Second, despite considerable development in signal-processing algorithms to identify transients in telescope images the problem remains complex, and will be exacerbated by the extremely large zoo of astrophysical transients that is expected in next-generation surveys such as LSST, which will enable the observation of up to one million objects per image. In this big data scenario, it is apparent that rudimentary applications of machine learning techniques that have complexity \(N \log \left(N\right)\), where \(N\) represents the number of points in a $d$-dimensional parameter space, will be too labor- and computational-intensive for the real-time validation of LSST detections, classifications and alerts~\cite{lsstbook,2019LSSTNat}. Machine learning techniques that fall into this category include nearest neighbor methods~\cite{Jones15679}, whose computational-intensive nature has motivated their implementation in  hardware architectures such as Graphic Processing Units (GPUs) and field-programmable gate arrays~\cite{nn_k_2010}.

In addition to these points, the existing cyber-infrastructure resources are oversubscribed~\cite{HPCUSE}, and MMA needs could only be accommodated by sacrificing other science applications in existing and planned supercomputer centers, as described in REF.\cite{NAP21886}. Therefore, a fundamental change is needed to secure the potential of MMA. To address these pressing issues requires the involvement of the astrophysics, data science, high performance computing (HPC), and cyber-infrastructure communities, and the creation of efficient communication channels between the GW astrophysics and time-domain astronomy communities to facilitate plans for joint, real-time observation campaigns. In this Expert Recommendation we make suggestions to accelerate the adoption of innovative signal-processing algorithms and computing approaches, driven by the big data revolution, that may help address computational challenges in MMA searches.

\vspace{4mm}
\noindent \textbf{Identification of optical counterparts}

\noindent The optical counterparts of BNS and NSBH mergers are known as kilonovae or macronovae~\cite{Metzger:2012}. Their emission spans the ultraviolet, optical, and near infrared bands, and encodes key information about ejected material powered by radioactive decays of r-process nuclei~\cite{SiegelMetzger3DBNS}.  

There are two issues related with the detection of such optical counterparts. First, the area, typically between a few tens to hundreds of square degrees~\cite{scenarioligo:2016LRR}, within which LIGO-type detectors can localize a MMA trigger includes many unrelated optical transients (for example, supernovae) whose properties are similar to those of optical counterparts of compact binary mergers. Second, optical counterparts evolve rapidly. Their fast decay rates (for example, the initial optical emission of the counterpart of GW170817 faded more than \(1\, \textrm{mag}\,\textrm{d}^{-1}\),  followed  by  a  longer-lived  red transient~\cite{drout:2017SCIENCE}) require identification within a few hours of the compact binary merger, with follow-up across the electromagnetic spectrum, especially with optical spectroscopy.

The location of a GW source source within the optical counterpart's host galaxy and the follow-up observations can shed light on the evolution of the progenitor system, its dynamics in the host galaxy, the production of heavy elements, and diagnostics regarding the mass and composition of the ejected material, properties of the circumstellar medium, and whether jets are generated during the event~\cite{Molley:2018Natur}. Optical counterpart identification requires a prompt response to the GW trigger to initiate a multi-filter imaging campaign  covering wide sky areas using large field-of-view, deep-imaging telescopes. Alternatively, smaller field-of-view telescopes can search for optical counterparts imaging pre-selected galaxies identified as plausible hosts. In both cases, these campaigns use automated discovery pipelines that perform image subtraction and artifact rejection. Deep Learning (DL) is already a standard component of such pipelines~\cite{igor:2017PASA}, and has a critical role in distinguishing astrophysical sources from noise. Rapid comparison with archive data will be valuable in pinpointing the optical counterparts. Furthermore, DL is now being used as a key discovery method in new pipelines, in the optical~\cite{seda:2018MNRAS} and at other frequencies. Future analyses may also use photometric classification methods, including DL, for the identification of transients~\cite{2010PASP..122.1415K,2018ApJ...857...51J}. The possibility of serendipitous kilonovae discovery using this approach in wide-field surveys such as LSST have been discussed in REFs\cite{2018ApJ...852L...3S,2018arXiv181210492S}.

GW sources and their optical counterparts may also be used to quantify how the Universe evolves. Assuming that Einstein's general relativity is the correct description of gravity, GW observations of compact binary mergers enable a direct measurement of the luminosity distance to their source~\cite{Schutz:1986Nature}. This property promotes them as standard-siren indicators, and may be used in conjunction with a catalog of potential host galaxies, which are accompanied with their associated redshifts, apparent magnitudes and sky localizations~\cite{darkgwss:2019}, to establish a redshift-distance relationship and measure the Hubble constant. GW standard measurements of the Hubble constant have already been carried out combining GW observations from the BNS merger GW170817 and its electromagnetic counterpart, which allowed an unambiguous identification of its host galaxy~\cite{2041-8205-848-2-L17,GWH:NaturA}. Multi-Messenger measurements of the Hubble constant are also possible with BBH mergers that have no electromagnetic counterparts so long GW observations are combined with photometric redshift catalogs. This has been demonstrated in the case of the BBH merger GW170814, and redshift information from the Dark Energy Survey Year 3 data~\cite{darkgwss:2019}. To ensure that galaxy catalogs are as complete as possible within the redshift range within which GW detectors observe BBH mergers~\cite{Maya:2018F}, DL algorithms~\cite{asad:2018K,Dom:2018MNRAS} are being used to classify galaxies in the Sloan Digital Sky Survey~\cite{SDSS:2011AJ} and the Dark Energy Survey~\cite{DES:2016MNRASOv} to construct galaxy catalogs at higher redshifts using the full 6-year data gathered by the Dark Energy Survey. These activities are essential to address similar challenges in future LSST-type surveys.

As the number of GW standard-siren measurements of the Hubble constant continue to increase in the near future, they may help understand whether the current tension between local measurements of the Hubble constant using type Ia supernovae observations~\cite{Riess:2019cxk}, and early universe measurements of the Hubble constant which combine cosmic microwave background data with the \(\Lambda\) cold dark matter model (where  \(\Lambda\) is the cosmological constant) of cosmology~\cite{Aghanim:2018eyx} is due to systematic errors or discrepancies that may usher in new discoveries~\cite{PhysRevLett.122.221301,Freedman:2017yms}.

\vspace{4mm}
\noindent \textbf{Real-time detection of GWs and neutrinos}

\noindent Real-time detection of GWs is essential to trigger time-sensitive searches to find their electromagnetic  and/or astro-particle counterparts. Because of the computationally-intensive nature and poor-scalablity of GW detection algorithms~\cite{HuertaBWOSG,2017Weitzel,HuertaES,Huerta:2017a,2016PhRvD..94b4012H}, the imminent increase in the number of GW observations thanks to the continuous increase in sensitivity of ground-based GW detectors, and the fact that available computing resources for these analyses is likely to remain the same, new solutions are needed to meet the increased demand for low-latency analyses. To harness the recent developments in data science that have revolutionized `big data' analyses in other science domains~\cite{2015Natur.521..436L}, DL prototypes have been developed for real-time detection and parameter estimation of non-spinning black holes on quasi-circular orbits that describe a 2D signal manifold (parametrized by the masses of the binary components \((m_1,\,m_2)\)), both in the context of simulated LIGO noise~\cite{geodf:2017a} and raw advanced LIGO noise~\cite{geodf:2017b}. These studies, which depended critically
on the use of the Blue Waters supercomputer for data generation, curation and for the training and testing
of DL models, have demonstrated that DL enables the detection of moderately eccentric, non-spinning BBH mergers and spin-precessing BBH mergers on quasi-circular orbits~\cite{geodf:2017a}, higher-order waveform modes of non-spinning BBHs on eccentric orbits~\cite{Rebei:2018R} and GW de-noising of true BBH mergers reported by the advanced LIGO and Virgo detectors~\cite{wei:2019W,Shen:ICASP2019,hshen:2017,shengeorge:PhysRevD97}. These studies have sparked the interest of the GW astrophysics community, leading to several developments in machine learning and DL for GW data analysis and source modeling~\cite{AlvinC:2018,2018GN,Nakano:2018,Dreissigacker:2019edy}.

Combining DL and HPC enables the design and training of neural network models using TB-size data sets of modeled waveforms, achieving state-of-the-art accuracy in actual detection scenarios. This approach  reduces the training stage from weeks to minutes, enabling detailed uncertainty quantification studies to assess the robustness of DL models. The first generation of neural network models designed at the interface of DL and HPC was introduced in REF.\cite{Shen:2019DLScale}, in which a template bank of over ten million modeled waveforms was used to train and validate a DL model within 10 hours. Furthermore, once fully trained, this DL algorithm may be used for real-time GW parameter estimation using a single GPU. This algorithm has been applied to estimate the astrophysical parameters of the catalog of BBH mergers reported in REF.\cite{o1o2catalog}, finding that deep learning results are consistent with those produced by Bayesian analyses. Future developments in this field concern the use of Bayesian neural networks to enable parameter estimation analyses endowed with full posterior distributions~\cite{NIPS2016_6117}. 

There is a pressing need to expand the scope of existing DL algorithms to cover the entire signal manifold that is available to existing GW detectors, and to enable real-time detection and characterization of BNS and NSBH mergers. These neural network models will have to be trained with significantly longer waveform signals. Since existing DL algorithms~\cite{geodf:2017b,Shen:2019DLScale} are capable of processing GW signals within milliseconds, one expects that even if DL algorithms for BNS observations are \(\sim1,000\) times slower than those used for BBH detection, one may still be able to extract BNS and NSBH signals from GW data in low-latency.

In addition to compact binary mergers, core-collapse supernovae~\cite{BurrowsCCSN,Burrows3d,Radice:2018usf,Mosta:2017geb} that occur within our Galaxy, may be observed in electromagnetic and gravitational waves and neutrinos~\cite{Janka:2016fox,Woosley:2006ie}. For core-collapse supernovae produced in the Large Magellanic Cloud, a neutrino burst may be observed though GWs may be too weak to be observed by second-generation GW detectors. For more distant core-collapse supernovae, only electromagnetic signatures may be observed~\cite{Gossan:2015xda}. 

As in the case of electromagnetic and gravitational observatories, DL has been explored for neutrino detection. DL is particularly suited for neutrino searches since neural networks are powerful feature extractors, and the core data challenge in neutrino detection concerns a correct categorization of high-level structures such as clusters, jets, tracks, showers and rings to tell apart signals from background noise. Recent studies provide evidence that convolutional neural networks may be used as event classifiers, achieving promising results separating signals from background noise~\cite{Aurisano:2016jvx}. Furthermore, graph neural networks provide an ideal framework to also take into account the topology of the detector to enhance neutrino detection. These studies showcase cutting-edge applications of computer vision to address contemporary challenges in MMA and high-energy physics~\cite{GNNIce}.

\vspace{4mm}
\noindent \textbf{Numerical relativity simulations}

\noindent MMA sources encompass the GW emission from BNS and NSBH mergers, the $\gamma$-ray emission from the compact object formed after merger, and an optical and infrared after-glow from the radioactive decay of r-process elements in ejected material. In the same way that numerical relativity simulations of BBH mergers have been critical for the detection and characterization of these sources with GW observations, numerical relativity simulations of BNS and NSBH mergers are critical to get insights into the physical processes that may lead to the production of electromagnetic and astro-particle counterparts, and to better interpret MMA observations~\cite{Hinderer:2018pei}. These modeling efforts do not currently benefit from DL, but recent studies have suggested the possibility to improve the efficiency and robustness of simulations, enabling the inclusion of detailed microphysics~\cite{Kim:DFluids,ling_kurzawski_templeton_2016,maulik_san_rasheed_vedula_2019,vigano:subgrid,PhysRevE.99.053113}, and a significance increase in the speed with which partial different equations are solved~\cite{berg2018unified,weinan2017deep}.

\begin{figure}[htb!]
\centering
\begin{subfigure}[t]{.45\textwidth}
\centering
\includegraphics[width=\linewidth]{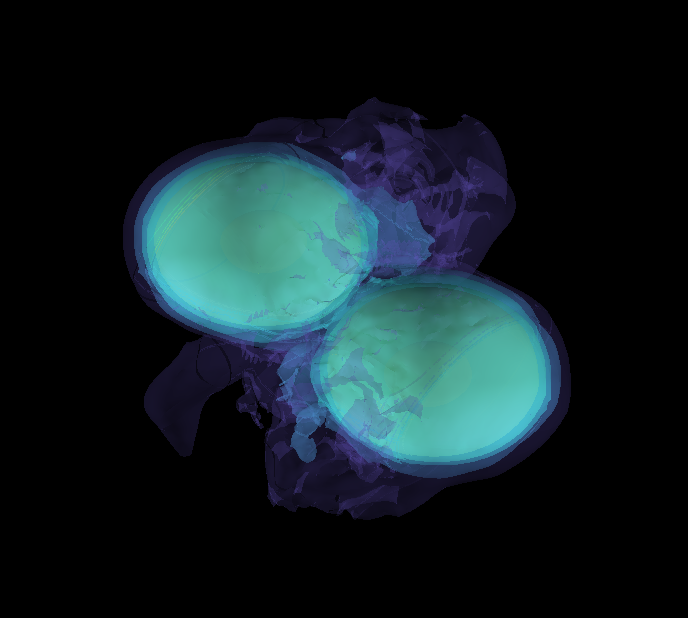} 
        \caption{}\label{fig:fig_a}
\end{subfigure}
\begin{subfigure}[t]{.45\textwidth}
\centering
\includegraphics[width=\linewidth]{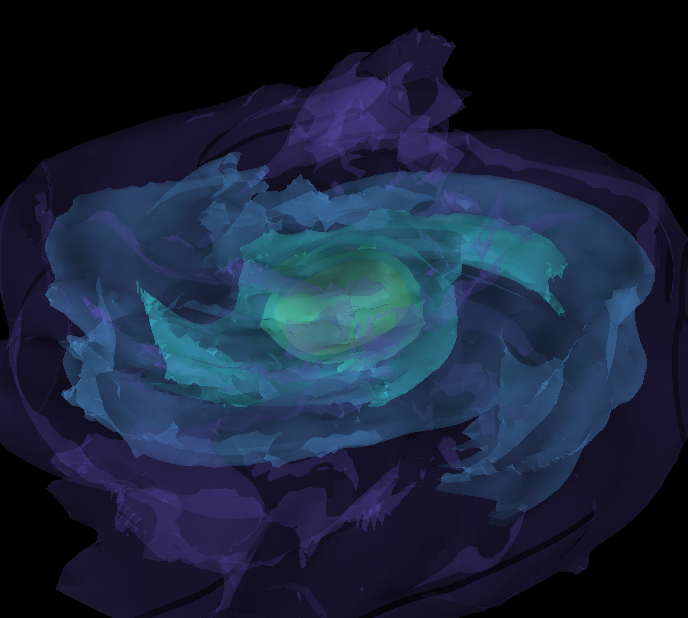}
\caption{}\label{fig:fig_b}
\end{subfigure}
\begin{minipage}[t]{\textwidth}
\caption{Visualization of the results of a numerical relativity simulation of two neutron stars before and after a merger. \textbf{a} Two neutron stars moving at a fraction of the speed of light shortly before they merge. Tidal forces start to tear apart the neutron stars and fling material away from them. Neutron star mergers may power short $\gamma$-ray bursts, as observationally confirmed with the binary neutron star merger GW170817 (REF.\cite{bnsdet:2017}). \textbf{b} The neutron star collision forms a hypermassive neutron star, which is surrounded by neutron-rich ejecta. The radioactive decay of the neutron-rich nuclei in the ejecta may produce a short-lived optical/near infrared weak supernova-like signal, known as kilonovae~\cite{Metzger:2012}. The observed features in GW170817 are broadly consistent with a kilonova model, indicating that r-process elements with a mass of \(0.05M_{\odot}\) were synthesized in that event~\cite{SiegelMetzger3DBNS,Kenta_18IJMPD}. This numerical relativity simulation was performed using the \texttt{Einstein Toolkit}~\cite{ETL:2012CQGra}, with the addition of the \texttt{WhiskyTHC}~\cite{Radice:2012cu} hydrodynamics code, and the \texttt{CTGamma} space-time evolution code. The physics in this simulation includes the fully General Relativity space-time evolution, a finite-temperature nuclear equation of state, and a neutrino evolution model. The simulation was performed and visualized on the Extreme Science and Engineering Discovery Environment (XSEDE) Stampede2 supercomputer.}
\end{minipage}
\end{figure}

Modeling the GW emission of MMA sources requires numerical relativity~\cite{DuezReview19} and the inclusion of magnetic fields to describe the formation of the jet and ejected material~\cite{BaiottiRezzollaReview}. Modeling the r-process elements requires neutrino transport and nuclear reaction networks~\cite{LippunerSkynet}. Whereas calculations with general relativistic magneto-hydrodynamics are now routine~\cite{prs15,Ruiz:2016rai}, the inclusion of more sophisticated neutrino and microphysics, is in its infancy. Using state-of-the-art numerical relativity software stacks, one can visualize the late inspiral phase of a BNS merger (FIG.~\ref{fig:fig_a}) in which the stars exchange matter due to tidal interactions. Upon merger, a hypermasssive neutron star is formed (FIG.~\ref{fig:fig_b}) and matter is ejected at high velocities. 

Accurate modeling of MMA sources relies on solving a realistic nuclear equation of state with the density, temperature and composition dependence, to capture not only the inspiral and merger effects, but, more importantly, to study the remnant with its subsequent black-hole-disk evolution~\cite{FernandezLongTermGRMHD,NouriPostMerger}. In the late merger phase, the inclusion of neutrino radiation transport and realistic magnetic fields are crucial to reliably simulate the short $\gamma$--ray burst and kilonova scenarios ~\cite{rad:2018ApJR}, whereas electromagnetic radiative transport is needed to provide realistic estimates of the disk geometry, mass accretion rate and electromagnetic luminosity and spectra~\cite{KasenOpacitiesBNS}. 

Astrophysics modeling involves a wide range of length and time scales. Often, the relevant scales can be modeled using mesh refinement~\cite{BERGER198964}. A promising application of machine learning is to model sub-grid physics that cannot currently be modeled from first principles at computational costs~\cite{chen2018neural} acceptable for parameter study simulations of turbulent motion during supernovae or neutron star collision~\cite{Radice:2015qva}. Sub-grid models tuned by a small number of high-resolution simulations avoid these costs~\cite{Giacomazzo:2014qba}.

\vspace{4mm}
\noindent \textbf{Cyber-infrastructure requirements}

\noindent Designing DL algorithms for real-time MMA analyses includes data curation, model training and inference, and result analysis. Training is the most computationally intensive step: it requires a significant amount of inter-process/thread communication and sequential calculations. Mature models can be trained using many CPUs/GPUs (horizontal scaling) with large batch sizes. New models, however, require a rounded balance of horizontal and vertical scaling, for example, accelerated CPU/GPU processing. In general inference is a parallel process: each input can be processed completely independently of any other input. Depending on the vertical scale of computation, the bottlenecks are often data retrieval, input/output latency, and bandwidth. Once fully trained, DL algorithms can process large datasets in real-time~\cite{geodf:2017a,geodf:2017b,Shen:2019DLScale}.

To enable real-time MMA analyses, the cyber-infrastructure must  support interactive access to large collections of disparate datasets, and the ability to apply DL algorithms to subsets of the data. These requirements suggest a low-latency interconnect, HPC-like system, with adequately provisioned data and computing resources.

At the time of writing, the Extreme Science and Engineering Discovery Environment (XSEDE) provides \(2.8\, \textrm{PFlop}\,\textrm{s}^{-1}\) of single precision GPU computing power through the Bridges and Comet clusters, both of which are nearing retirement. New resources are expected to become available on the 2020 time frame, but it is unknown what the configuration of these resources will be, and whether they will effectively support the MMA computing needs. The National Science Foundation (NSF)-funded Frontera supercomputer will provide only \(8\, \textrm{PFlop}\,\textrm{s}^{-1}\) of 32-bit precision computational power through a small GPU section of the system, which will be readily consumed by basic MMA machine learning applications. Other NSF-funded investments include the DL project at the National Center for Supercomputing Applications, which consists of 16 IBM Power9 servers with four NVIDIA V100 GPUs, and which supports \texttt{TensorFlow} and other DL frameworks in a container-based environment. The Department of Energy supercomputing platforms Summit and Titan at Oak Ridge National Laboratory are equipped with over 27,000 NVIDIA Volta V100 GPUs and over 18,000 K20X GPUs, respectively. Piz Daint, Europe's fastest supercomputer, is equipped with 5,320 hybrid compute nodes, each equipped with two Intel Xeon E5-2690 v3 processors and a single NVIDIA Tesla P100. The system also has 1,431 GPU-less XC40 nodes, each of which has two Intel Xeon E5-2695 v4 processors. According to EuroHPC, a joint initiative between the European Union and European countries to spearhead innovation for the European HPC ecosystem, the computation capabilities in the European Union do not match the current demand for computing and data needs from the academic and industry sectors in Europe~\cite{eurohpc}. Furthermore, the fastest European supercomputers depend on non-European technology. To address these problems, and develop a strategic plan for innovation and competitiveness, EuroHPC aims to buy and install two pre-exascale machines by 2020, and 3-4 additional petascale machines. It is expected that by 2023 Europe will have two exascale machines and one post-exascale machine that may support the first hybrid HPC/Quantum computing infrastructure in Europe.

Across the spectrum of computational sciences, the supply of large-scale computing resources does not meet the demand~\cite{HPCUSE,NAP21886,eurohpc}. Funding-agency research grants for computational projects support people, but often do not guarantee computer resources. Deployment and operation of sufficient agency resources to cover the majority of the predicted demand over the next few years might best be served by organizing around individual resources scaled around 50-100 PFlops/s double precision, 10-100 petabytes storage, and external network data intake capacity sufficient for the analysis of the expected observational data. The estimated demand will likely require multiple instances of resources/centers at this scale. These should provide common HPC, machine learning/DL, and data analytics software environments.

The landscape of existing and planned NSF-funded HPC cyber-infrastructure indicates a pressing need to expand these resources to meet the MMA needs~\cite{HuertaES,8109152}. Given the much broader applicability of machine learning/DL approaches, the agencies should invest in the resources and services necessary to enable the success of MMA and other community efforts reliant on these approaches. These resources include appropriate storage and data-sharing environments that can support multidisciplinary collaborations, along with an adequate networking infrastructure to support both bulk data transport and rapid notifications. We also advocate for the exploration of newer technologies, such as, tensor processing units, quantum computers, and  their applicability to these challenges.

\vspace{4mm}
\noindent \textbf{Future needs of MMA} 

\noindent MMA heavily relies on sharing real-time alerts among various instruments. These alerts would be prioritized by every instrument, given their observing capabilities and science goals. A common tool to help identify potential host galaxies of GW sources is highly desirable~\cite{Arcavi:2017vbi}, as well as coordination between groups to determine which sources may be optimally followed-up with wide-field, all-sky survey instruments, or through target of opportunity observations with narrow field of view telescopes where the number of observable sources is significantly limited~\cite{Coughlin:2018lta}. DL classifiers could automate these time sensitive tasks in existing services such as the NASA/IPAC Extragalactic Database (NED) Gravitational Wave Follow-up (GWF) Service~\cite{nednasa} and the Astrophysical Multi-Messenger Observatory Network (AMON)~\cite{amoncite}.

The engagement of the next generation of telescopes (LSST, Thirty Meter Telescope, Extremely Large Telescope, and so on) in MMA should be defined, as should observing policies for existing large telescopes: the observing strategy, mechanism and software infrastructure should be flexible enough to allow a rapid reaction to alerts regarding important potential targets. Pre-defining the amount of time and resources each next generation telescope and survey will spend on MMA follow-ups, involving the transient and static sky science communities, would avoid potential conflicts and waste of resources~\cite{Marshall:2017wph,keck_plan,lsstbook,Cowperthwaite:2019zqg}. Archives for large telescopes should ideally support real-time ingestion of freshly-acquired data and, and as much as possible, make MMA follow-up data public immediately after receipt.

Ideally, a common, public database would be build to centralize all information available for every recorded event. This will minimize the duplication of common tasks (such as data collection, galaxy cross-matching, ranking, and so on) among follow-up teams. The centralized database could be coupled with classification algorithms and catalogs cross-matching. In this sense, the infrastructure of the LSST data management system and the alert brokers under development  are good examples~\cite{Narayan:2018rxv,Smith_2019}. The automation of telescopes is an important part of maximizing the number of observed MMA events (for example the Astronomical Event Observatory Network program will provide programmable access to a number of telescopes~\cite{aeon}).

Finally, it is crucial that funding agencies explore the best way to manage competitive proposals pursuing the same scientific goals. Given the scientific value of some MMA events, several teams are likely to simultaneously request the same observations from the same facilities, creating unneeded duplication and the potential for conflict. Although competition can be valuable on a scientific level, on the infrastructure level it often leads to inefficiency. Given the increasing numbers of MMA events available thanks to improving instrument sensitivities, this inefficiency could become a significant drain on the limited observing resources. Whereas different approaches may be used to manage the technical issues relevant to these problems, the scientific community may still need to reach consensus on the common requirements and science priorities (for instance within the Astro2020: Decadal Survey on Astronomy and Astrophysics~\cite{astro2020}) that would allow the funding agencies to plan a common infrastructure and provide guidance to both the review and decision process regarding competing usage proposals.

\vspace{4mm}
\noindent \textbf{Community building}

\noindent By the very nature of the field, the science goals of MMA cannot be realized by any one individual or small team. Community efforts are essential to building the necessary software to manage the needs of large-scale collaborations~\cite{8565942}. Successful examples in astronomy (for example the \texttt{Astropy} community project, the \texttt{yt} community project, and the LSST-Dark Energy Science Collaboration) can act as models for building an MMA software community. 

Other data-intensive scientific domains, such as high-energy physics (HEP), share similar challenges in building and sustaining a cohesive community, workforce training, development and advancement, access and delivery of large amounts of data, low-latency data analysis, and using machine learning techniques. In HEP, 18 workshops in two years engaged key national and international partners from HEP, computer science, industry, and data science to generate over eight community position papers, including a software institute Strategic Plan~\cite{S2I2HEP-SPj} and a Community White paper~\cite{CWP} as a roadmap for HEP software and computing research and development over the next decade. The MMA community should consider what can be learned from this intensive community effort, and use the transferable solutions.

A similar effort was already set in motion with the NSF-funded project Community planning for Scalable Cyber-infrastructure to support Multi-Messenger Astrophysics (SCiMMA)\cite{whitepaper:SCIMMA}, which seeks to identify the key questions and cyber-infrastructure projects required by the MMA community to take full advantage of current  and imminent next-generation facilities.
SCiMMA held a number of workshops to bring community members together, and is now writing white papers on various sub-topics, including: systems development; data management, communication, and collaboration; analysis, inference, and machine learning; modeling and theory; education, training, and workforce development; and management. SCiMMA will also develop a strategic plan for a scalable cyber-infrastructure institute for MMA laying out its proposed mission, identifying the highest priority areas for cyberinfrastructure research and development for the US-based MMA community, and presenting a strategy for managing and evolving a set of services for the broader community.

\vspace{4mm}
\noindent \textbf{Conclusions}

\noindent For the challenges discussed in the previous section we make a number of recommendations summarized in Table~\ref{tab:rec_all}. We identified two pillars that require immediate attention to fully realize the MMA science program: first, increase the speed and depth of signal processing algorithms for real-time detection of GWs and their astro-particle and electromagnetic counterparts; and second, identify the cyber-infrastructure resources to simulate and search for MMA events in the ever increasing and disparate data sets. Furthermore,  DL methods to accelerate the solution of partial differential equations could be used to greatly increase the speed, accuracy and robustness of numerical relativity simulations of MMA sources~\cite{weinan2017deep,berg2018unified}.

There is concern regarding the availability of future cyber-infrastructure facilities for MMA data analytics. Furthermore, whereas DOE HPC platforms are particularly suited for DL analytics, they are highly oversubscribed. Therefore, we recommend reaching out to diverse advanced cyber-infrastructure centers in the US (Summit, Sierra, Frontera, Bridges-2, Theta, Lassen, etc.) and Europe (Piz Daint, SuperMUC-NG, etc.) to obtain computational resources for DL at scale. This is critical because the convergence of DL with HPC is essential to train deeper and more accurate neural network models with TB-size training data sets. This approach is essential to ensure that DL algorithms characterize the sources' parameter space that is available to observatories. Once these models are fully trained, they can be used for real-time MMA searches using minimal computational resources. 

We also recognize the need for policy making regarding data acquisition and  data sharing at astronomical observatories and for facilitating the interaction among the different MMA sub-fields. The MMA community should put in place mechanisms to reward its members, allow them to develop and provide career paths for its members beyond academia. We suggest collaborating with industry partners who, in addition to co-funding data science workshops and bootcamps for MMA researchers, could also recruit members of this community.

\begin{table}
\begin{center}
    \begin{tabular}{ | p{2.8cm} | p{13.8cm} |}
    \hline
    \multicolumn{1}{|c|}{Area} & \multicolumn{1}{c|}{Recommendation} \\ \hline
    Numerical relativity & 
    \noindent Improve coupling and cross-talk between numerical relativity and modeling tools and build a pipeline for predicting observables\newline
    \noindent Accelerate community efforts to develop and release open source versions of modeling codes, especially microphysics and transport packages\newline
    \noindent Explore publicly funded opportunities to develop numerical relativity software critical for MMA interpretations such as through the NSF or the DOE office of science \\ \hline
    Gravitational wave astrophysics and neutrino physics& 
    \noindent Accelerate the convergence of DL and HPC to reduce training stage from weeks to minutes of DL models that cover the complete signal manifold of black hole mergers \newline
    \noindent Design Bayesian neural network models for GW parameter estimation covering the full signal manifold of black hole mergers \newline
    \noindent Design neural network models for real-time detection of compact binaries involving neutron stars and core-collapse supernovae\newline
    \noindent Design and deploy production-scale neural network classifiers to categorize particle interactions in neutrino detectors\newline
    \noindent Accelerate the development of graph neural networks that are tailored for the specific topology of existing detectors \\ \hline
    Electromagnetic surveys & 
    \noindent Continue the development and adoption of software, including the robotization of telescopes, to facilitate the rapid, automated, coordinated response to GW triggers from multiple telescopes observing across the EM spectrum and leverage machine learning for decision making and optimization \newline
    \noindent Continue the development of algorithms and software for the classification and identification of EM counterparts of GW events, including the integration of DL in event selection and classification from images and hybrid data sources (direct integration of images and catalog data) \newline
    \noindent Continue the development of simulations and emulations in astrophysics modeling to fully explore the parameter space of the EM counterparts of GW events  \newline
    \noindent Create and facilitate automated access to comprehensive catalogs of archival and new observations to be used in event selection and classification
    \newline
    \noindent Create communication channels between concurrently observing facilities to share plans for observing targets
    \\\hline
    Cyber-infrastructure for DL & 
    \noindent Develop public training and test data sets, and a mock system infrastructure to test trained models\newline
    \noindent Funding agencies (including NSF and DOE in the US) should agree on a split of responsibility for the common cyber-infrastructure needed to support all projects and researchers, rather than individual projects \newline 
    \noindent Funding agencies should ensure that the majority of their own computational research funding also covers the proposed computing resources (through their own systems and inter-agency agreements)\newline
    \noindent Maintaining a current and consistent DL and workflow software stack are essential computational resources for theoretical research and practical applications at scale, and provide critical tools for community training and education\newline
    \noindent Development of open-source software products and scalable and efficient DL algorithms which maximize use of HPC resources are needed\newline
    \noindent  Integrating scalable machine learning with developments in computational methods and HPC platforms has the potential to benefit many key areas of interest to funding agencies.\\\hline
    Community building & 
    \noindent Develop trust between projects, individual researchers, and these two  groups; then formalize that trust into agreements about data collection, data access, credit and authorship, and so on. \newline
    \noindent Build on the example of successfully-governed, open source community projects to develop community maintained software systems  \newline
    \noindent Organize workshops to train users and encourage participation \newline 
    \noindent Develop mechanisms to provide incentives (opportunities to represent scientific communities at conferences with invited talks and keynotes), rewards (including fellowships, and support to attend and participate in prestigious summer and winter schools) and recognition (having leadership roles in scientific communities) for success  
    \newline
    \noindent Develop rules and guidelines for data  sharing and publication to assure ethical behavior and avoid waste of EM observational resources
    \\ 
    \hline
    \end{tabular}
\end{center}
\caption{Summary of recommendations to realize the science goals of Multi-Messenger Astrophysics. Notes: NSF, National Science Foundation; DOE, Department of Energy; HPC, High Performance Computing; GW, gravitational waves; EM, electromagnetic; DL, deep learning}
\label{tab:rec_all}
\end{table}
\newpage

\bibliography{references,ref_two}

\vspace{4mm}
\noindent \textbf{Acknowledgements}\\
The authors gratefully acknowledge support from NVIDIA, Argonne Leadership Computing Facility, Oak Ridge Leadership Computing Facility, and the National Science Foundation through grant NSF-1848815. Artwork in this manuscript was supported in part by the National Science Foundation through grants ACI-1238993, NSF-1550514 and TG-PHY160053.

\vspace{4mm}
\noindent \textbf{Author contributions}\\
E.A.H. led and coordinated the writing of this Expert Recommendation. All authors contributed to developing the ideas, and writing and reviewing this manuscript. S.R. produced the artwork in figure 1.

\vspace{4mm}
\noindent \textbf{Competing interests}\\
The authors declare no competing interests. 

\vspace{4mm}
\noindent \textbf{Reviewer information}\\
\textit{Nature Reviews Physics} thanks Brant Robertson, Viviana Acquaviva and the other, anonymous, reviewer(s) for their contribution to the peer review of this work.

\vspace{4mm}
\noindent \textbf{Publisher's note}\\
Springer Nature remains neutral with regard to jurisdictional claims in published maps and institutional affiliations.

\end{document}